\documentclass[12pt,twoside]{elsart}
\usepackage{amsfonts}
\usepackage{epsfig}
\usepackage{fancyhdr}
\usepackage[latin1]{inputenc}
\usepackage{natbibf}
\usepackage{fancyhdr}
\usepackage{amsmath}
\usepackage{graphicx}
\usepackage{hyperref}
\usepackage{amssymb}
\usepackage{dsfont}
\newtheorem{theorem}{Théorème}[section]
\newtheorem{definition}{Définition}[section]
\newtheorem{property}{Propriété}[section]

\newtheorem{proposition}{Proposition}[section]
\newtheorem{remark}{Remark}[section]
\numberwithin{equation}{section}

\renewcommand{\cite}{\citet}
\usepackage{amssymb}
\begin{document}
\begin{frontmatter}
\title{Régularisation de l'équation de Langevin en dimension
$1$ par le mouvement Brownien fractionnaire}
\author[k]{Lounis Tewfik} et \author[b]{Saïd Bouabdellah}
\address[k]{tewfik.lounis@gmail.com }
\address[b]{said.bouabdellah@gmail.com}
\begin{abstract}
The main goal of this paper is to provide a fractional stochastic
differential equation modelling the physical phenomena governed by
the Langevin equation in 1-dimension. A generalized equation
leaning on the fractional Brownian motion (fBm) will be proposed,
the later will allow a description of the complexity of the
physical systems which escape any prediction of the of the
standard Langevin equation. We shall begin at first to remind the
basic notions of the standard Brownian motion (Bm) and the
fractional Brownian motion (fBm), then, we shall establish a
generalization to long memory of the Langevin equation.\vskip7pt
\noindent {\bf Résumé.}\\
\vskip7pt \noindent L'objectif de ce travail est de donner une
équation différentielle stochastique fractionnaire modélisant les
phénomènes physiques gouvernés par l'équation de Langevin en
dimension $1$. Une équation généralisée s'appuyant sur le
mouvement brownien fractionnaire mBf sera proposée, elle permettra
une description de la complexité des systèmes physiques qui
échappent à toute prédiction de l'equation de Langevin standard.
On commencera dans un premier temps par rappeler les notions de
base du mouvement brownien standard mB et le mouvement brownien
fractionnaire mBf, puis, nous établirons une généralisation à
mémoire longue de l'équation de Langevin. Ce travail s'inscrit
dans le cadre de la continuité de l'application du travail de
\cite{TL} inspiré des travaux de  \cite{DN}.
\end{abstract}
\end{frontmatter}
%\section{Introduction}
\section{La mémoire courte et longue}
Nous connaissons de nombreux modèles stochastiques à mémoire
courte: on peut citer les variables indépendantes, les variables
m-indépendantes, certain processus de Markov, les moyennes
mobiles, la plupart des processus autorégressifs à moyenne mobile
(ARMA) et de nombreux processus linéaires. L'avantage de la courte
mémoire est qu'elle induit, le plus souvent, de nombreux théorèmes
limites comme les lois des grands nombres, des théorèmes centraux
fonctionnels et des théorèmes de grandes déviations. Lorsque la
mémoire est longue, la situation est souvent différente. Il y a au
moins un siècle que les astronomes ont noté l'existence de séries
d'observations à mémoire persistante. Depuis lors, des phénomènes
de même type ont été observés, notamment, en chimie, en
hydrologie, en climatologie, et en économie. De telles séries
posent évidemment d'intéressants problèmes statistiques et leur
modélisation ainsi que leur traitement statistique ont déjà fait
l'objet de plusieurs travaux de recherches. Dans \cite{JB}, on
trouve quelques exemples historiques traités par modélisation
brownienne fractionnaire. La modélisation de certains phénomènes
physiques, dits à mémoire longue, peut s'effectuer par
l'introduction de termes intégro-différentiels à noyaux faiblement
singuliers (càd localement intégrables, mais pas nécessairement
continus) dans les équations de Hamilon - Jacabi ou Euler -
Lagrange  qui régissent la dynamique physique. Il existe également
des applications à la modélisation en chimie des polymères ou à la
modélisation de la dynamique à l'interface de structures
fractales: voir \cite{LEM} pour l'aspect physique appliquée et,
par exemple, \cite{MG} pour l'aspect physique théorique. De plus,
la dérivation fractionnaire peut paraître naturellement lorsqu'un
phénomène dynamique est fortement conditionné par la géométrie du
problème: un exemple simple, très instructif, est présenté dans la
référence \cite{TOR}.
\section{Préliminaires}Dans toute cette
partie, $(\Omega,\mathcal{F} ,P)$ est un espace de
probabilit\'{e}, $(\mathcal{F}_t)_{t\geqslant0}$ une filtration
sur cette espace, $\{X_t,t\geqslant0\}$ un processus adapt\'{e}
\`{a} valeurs dans $ R^d$, $(d\geqslant1)$, $\{B_t ,t\geqslant0\}$
est $\mathcal{F}_t$ mouvement brownien. $[0,T]$ est un intervalle
borné dans $\mathbb{R}$ sur lequel on d\'{e}finit une subdivision
$(t_i)$ o\`{u} $i\in\{1,\dots,n\}$ et $t_0=0<t_1<\dots<t_n=T$, et
on pose $\Delta=\max(t_i-t_{i-1})$.
\subsection{Mouvement brownien ordinaire}
Un mouvement brownien ordinaire $B=\{B_t, t\geqslant 0\}$ (ou
processus de Wiener ) est un processus sur $(\Omega,{\mathcal F},
{\mathcal F}_t, P)$ adapt\'{e}, continu et admettant les
propri\'{e}t\'{e}s suivantes~:
\begin{itemize}
\item $P(B_0 = 0 )=1$ $\mathbb{P}$-p.s sur $\Omega$. \item
$\forall s\geqslant0$ et $\forall t \geqslant 0$ avec $t \geqslant
s$; $(B_t - B_s)$ est une variable r\'{e}elle, de loi gaussienne,
centr\'{e}e et de variance $t-s$. \item $\forall s\geqslant0$ et
$\forall t \geqslant 0$ avec $t \geqslant s$; $(B_t - B_s)$ est
ind\'{e}pendant de $\mathcal{F}_u$ o\`{u} $u\leqslant s$.
\end{itemize}
\subsection{Quelques propri\'{e}t\'{e}s du mouvement brownien
standard}
\subsubsection{Variation quadratique du mouvement brownien standard}
\begin{definition}
La variation quadratique du mouvement brownien \\$B=\{B_t,
t\geqslant0\}$ est la limite dans $L^2(\Omega)$ de
$\sum_{i=1}^n\vert B_{t_i}-B_{t_{i-1}}\vert^2$ quand $\Delta$ tend
vers 0.
\end{definition}
\begin{theorem}
La limite ci-dessus existe dans $L^2(\Omega)$ et vaut $T$.
\end{theorem}
\subsubsection{Propri\'{e}t\'{e} de Markov}
\begin{definition}
La fonction al\'{e}atoire $X=\{X_t, t\geqslant0\}$ est un
processus de Markov par rapport \`{a} $\mathcal{F}$, si pour
$s\geqslant0, t\geqslant0$ et tout bor\'{e}lien $A$ de
$\mathbb{R}^d$ on a:
\begin{equation}
P\left(X_{t+s}\in A\vert\mathcal{F}_t\right)=P\left(X_{t+s}\in
A\vert X_{t}\right).
\end{equation}
\end{definition}
Cette derni\`{e}re \'{e}galit\'{e} s'appelle propri\'{e}t\'{e} de
Markov, cela signifie que la loi de $X_{t+s}$ sachant
$\mathcal{F}_t$ ne d\'{e}pend que de $X_t$. "La loi du futur ne
d\'{e}pend du pass\'{e} que par le pr\'{e}sent".
\begin{remark}
Pour plus de détails sur les démonstrations, on peut se rapporter
au livre de  \cite{TM}.
\end{remark}
\section{Mouvement brownien fractionnaire}
Un mouvement brownien fractionnaire $B^H=\{B^H_t, t\geqslant0\}$,
de paramètre $H$ dans $(0,1)$, est un processus gaussien
centr\'{e} v\'{e}rifiant les conditions suivantes~:

\begin{enumerate}
 \item $B^H$ est un processus stationnaire;
 \item $\mathbb{E}(B^H_t)=t^{2H}$;
 \item $B^H_0=0$.
\end{enumerate}
Le param\`{e}tre $H$ est l'indice de Hurst.\\
\begin{remark}
\begin{enumerate}
    \item Si $H=\frac{1}{2}$, $B^H$ est le
mouvement brownien standard.
    \item Si $H\neq\frac{1}{2}$, les accroissements du mBf ne
    sont pas indépendants.
\end{enumerate}
\end{remark}
\subsection{Quelques propri\'{e}t\'{e}s du mBf}
\begin{property}
 $B^H$ admet comme fonction de covariance, la
fonction $R_H$ d\'{e}finie pour tout
$(s,t)\in\mathbb{R}^+\times\mathbb{R}^+$ par:
\begin{equation}
 R_H(t,s)=\frac{1}{2}(t^{2H}+s^{2H}-\vert t-s\vert^{2H}).
\end{equation}
\end{property}
\begin{remark}
Si $H=\frac{1}{2}$, alors:
\begin{eqnarray*}
 \mathbb{E}(B^H_t B^H_s) &=& \frac{1}{2}\big(t+ s-\vert t-s\vert\big) \\
 &=& s \\
&=& s\wedge t\\ &=& \textrm{Cov}(B_t,B_s).
\end{eqnarray*}
\end{remark}
On retrouve ainsi la covariance du movement brownien standard.
\begin{property}
La fonction de covariance $R_H$ d\'{e}finie de
$\mathbb{R}^+\times\mathbb{R}^+$ dans $\mathbb{R}$ est continue,
semi-d\'{e}finie positive et sym\'{e}trique.
\end{property}
\subsection{Correlations à long terme}
Le mouvement brownien fractionnaire est caractérisé par la
présence de corrélations à long terme lorsque $H$ est différent de
$\frac{1}{2}$. Un exposant supérieur à $\frac{1}{2}$ révèle des
phénomènes de persistance, c'est-à-dire que l'évolution du
processus tend à suivre ses tendances. Si le processus a augmenté
précédemment, la probabilité est forte qu'il continue à le faire.
Les processus persistants ont une mémoire à long terme,
c'est-à-dire qu'il existe une corrélation à long terme entre les
événements actuels et les événements futurs. Chaque observation
porte la mémoire des événements qui l'ont précédé. \`{A}
l'inverse, un exposant inférieur à $\frac{1}{2}$ révèle un
phénomène d'anti-persistance, dans ce cas les accroissements
successifs tendent à être négativement corrélés. Une augmentation
de la variable tend à être suivi d'une diminution, et inversement.
\section{Représentation d'un mBf sur un intervalle\label{D.N}}
Dans cette partie, la représentation du mBf sera faite sur un
intervalle $[0,T]$, une technique qui s'appuie sur les noyaux de
carré integrable. Fixons l'intervalle $[0,T]$ et considérons les
deux cas: $H>\frac{1}{2}$ et $H<\frac{1}{2}$. \vspace{3mm}
\textbf{{Premier cas: $H>\frac{1}{2}$}} \vspace{3mm} \cite{DN}
avait prouvé que:
\begin{equation}
   B^H_t=\int_0^tK_H(t,s)dB_s
\end{equation}
Où la fonction de carré intégrable $K_H$ définit par:
\begin{equation}
K_H(t,s)=c_Hs^{\frac{1}{2}-H}\int_s^t(u-s)^{H-\frac{3}{2}}u^{H-\frac{1}{2}}du,
\end{equation}
avec $t>s$, et en choisissant $c_H=\Big[\frac{H(2H-1)}
{\beta(2-2H,H-\frac{1}{2})}\Big]^{\frac{1}{2}}$, on a:
\begin{equation}
 \int_0^{t\wedge s}K_H(t,u)K_H(s,u)du = R_H(t,s).
\end{equation}
Notons que:
\begin{equation}
\frac{\partial K_H}{\partial
t}(t,s)=c_H\left(\frac{t}{s}\right)^{H-\frac{1}{2}}(t-s)^{H-\frac{3}{2}}.
\end{equation}
\vspace{3mm} \textbf{{Deuxième cas: $H<\frac{1}{2}$}} \vspace{3mm}
 \cite{DN} avait prouvé que:
\begin{equation}
B^H_t=\int_0^tK_H(t,s)dB_s.
\end{equation}
Où la fonction de carré integrable $K_H$ définit par:
\begin{equation}
K_H(t,s)=c_H\left[\left(\frac{t}{s}\right)^{H-\frac{1}{2}}(t-s)^{H-\frac{1}{2}}-
   (H-\frac{1}{2})s^{\frac{1}{2}-H}
   \int_s^tu^{H-\frac{3}{2}}(u-s)^{H-\frac{1}{2}}du\right],
\end{equation}
avec $t>s$, et en choisissant $c_H=\Big[\frac{2H}
{(1-2H)\beta(1-2H,H+\frac{1}{2})}\Big]^{\frac{1}{2}}$, on a:
\begin{equation}
 \int_0^{t\wedge s}K_H(t,u)K_H(s,u)du = R_H(t,s)
\end{equation}
Notons que:
\begin{equation}
\frac{\partial K_H}{\partial
t}(t,s)=c_H(H-\frac{1}{2})\left(\frac{t}{s}\right)^{H-\frac{1}{2}}(t-s)^{H-\frac{3}{2}}.
\end{equation}
\section{Principe d'invariance}
Soit $\{\xi_i, i\geqslant 1\}$ une suite de variables aléatoires
réelles indépendantes et de même loi, avec $\mathbb{E}\xi=0$ et
$\mathbb{E}\xi^2=\sigma^2$. On considère la marche aléatoire:
\begin{equation}
S_n=\xi_1+\dots+\xi_n\quad,\quad n\geqslant 1.
\end{equation}
Le théorème de la limite centrale implique que
$\frac{S_n}{\sigma\sqrt{n}}$ converge en loi vers
$\mathcal{N}(0,1)$ quand $n\rightarrow \infty$. Derrière cette
convergence de variables aléatoires se cache une convergence de
fonctions aléatoires. Définissons la ligne polygonale $X^n$
extrapolant la marche aléatoire $S$:
\begin{equation}
X^n(t)=\frac{\sum_{i=1}^{[nt]}\xi_i+(nt-[nt])\xi_{[nt]+1}}{\sigma\sqrt{n}}\quad,
\,\quad t\in\mathbb{R}_{+}
\end{equation}
$X^n$ est une fonction aléatoire réelle continue. On calcule
facilement ses accroissements:
\begin{eqnarray*}
\begin{pmatrix}
  X^n(t_1) \\
  X^n(t_2)-X^n(t_1) \\
\end{pmatrix}&=&\frac{1}{\sigma\sqrt{n}}\begin{pmatrix}
  \sum_{i=1}^{[nt_1]}{\xi_i} \\
  \sum_{i=[nt_1]+1}^{[nt_2]-[nt_1]}{\xi_i}  \\
\end{pmatrix}
+o_{L^2}(1)\\&\underrightarrow{\textrm{loi}}
&\mathcal{N}(0,t_1)\otimes\mathcal{N}(0,t_2-t_1)
\end{eqnarray*}
si $n\rightarrow \infty$, par la théorème de la limite centrale.
Plus généralement, pour une subdivision arbitraire
$0=t_1<t_2<\dots <t_m$, on a:
\begin{equation}
\lim_{n\rightarrow
\infty}[X^n(t_i))_{i=1}^{m}-X^n(t_{i-1})]_{i=1}^{m}=[B(t_i)-B(t_{i-1})]_{i=1}^{m}
\end{equation}
en loi, et donc la convergence en loi des martingales
fini-dimensionelles:
\begin{equation}
\lim_{n\rightarrow\infty}[X^n(t_i)]_{i=1}^{m}=[B(t_i)]_{i=1}^{m}.
\end{equation}
On admet le théorème suivant:
\begin{theorem}[Donsker (1951): Principe d'invariance]La suite de
processus aléatoires $\{X^n, n\geqslant1\}$ converge en loi vers
le mouvement brownien. En d'autres termes, soit $P_n$ la loi de
$X^n$ sur $\mathcal{C}(\mathbb{R}_+,\mathbb{R})$, et $P_B$ la loi
de B. Alors,$$P_n\rightarrow P_B \quad\text{étroitement}$$.
\end{theorem}
\section{\'{E}quation de Langevin}
On considère le modèle d'une particule de masse $m$ qui se déplace
sur une droite, soumise à une force de friction et une force
dynamique extérieure $F_t$ qui modélise les chocs avec les autres
particules. On suppose que la particule n'est soumise à aucune
autre force extérieure. D'après la seconde loi de la dynamique non
relativiste de Newton dans un référentiel Galillien, la position
$x_t$ est alors solution de l'équation différentielle:
\begin{equation}
mx_t''=-bx'_t+F_t,
\end{equation}
qu'on peut écrire aussi sous la forme:
\begin{equation}
mv'_t=-bv_t+F_t.\label{Eq. Langevien}
\end{equation}
avec $b>0$, le coefficient de friction et $x_t$ la position
instantanée de la particule. Si la force extérieure $F_t$ est dû à
des chocs, nombreux et petits, le principe d'invariance de la
section précédente nous incite à considérer que
$\int_0^t{F_sds}=\sigma B_t$, pour un coefficient $\sigma
>0$, paramétrisant l'intensité de chocs et leurs amplitudes. La force
$F$ est alors un bruit blanc i.e. la << dérivée >> d'un mouvement
brownien, et l'équation ci-dessus est improprement écrite, elle
doit porter sur les différentielles stochastiques. La vitesse
$V_t=x'_t$ est solution de l'équation différentielle stochastique:
\begin{equation}\label{E.L}
mdV_t=-bV_tdt+\sigma dB_t
\end{equation}
au sens  où, presque sûrement:
\begin{equation}
V_t-V_0+\int_0^t{\frac{b}{m}V(s)ds}=\frac{\sigma}{m}
B_t\quad,\quad \forall{t}\geqslant0.
\end{equation}
Cette équation différentielle stochastique fut proposé par Paul
Langevin en 1908.
\subsection{\'{E}quation de Langevin en dimension 1}
Soit $B$ un $(\mathcal{F}_t)_{t\geqslant0}$ mouvement brownien
unidimensionnel. On admet la proposition suivante:
\begin{proposition} La solution de l'équation de Langevin (\ref{E.L}), partant de $V_0\in L^2(\mathcal{F}_0)$, est donnée par le
processus d'\textbf{Ornstein-Uhlenbeck}, défini par:
\begin{equation}
V_t=e^{-\frac{b}{m}t}V_0+\int_0^t{e^{-b(t-s)}\frac{\sigma}{m}
dB_s}.
\end{equation}
\end{proposition}
On peut montrer par un simple calcul que:
\begin{equation}
\label{esperance} \mathbb{E}V_t=e^{-\frac{b}{m}t}\mathbb{E}V_0.
\end{equation}
\subsection{mBf et l'équation de Langevin }
Si on examine de prêt l'équation de Langevin, on est parti d'une
force aléatoire $F_t$ qui est approximé par un bruit blanc. Vu que
le mouvement brownien standard est un processus de markov (par
rapport à sa filtration), et pour mieux mettre en évidence
l'influence du passé, il sera judicieux d'élaborer un modèle qui
tient compte de l'effet du passé tout en gardant les caractères
gaussiens et la stationnarité. \`{A} cet effet, on introduit le
mouvement brownien fractionnaire (processus non markovien pour
$H\neq\frac{1}{2}$). \vspace{3mm} \cite{strook} avait montré qu'on
peut approcher un bruit blanc par un processus
$(\theta_\epsilon)_{\epsilon
>0}$. Quand $\epsilon\rightarrow 0$ cette convergence est en loi
dans $\mathcal{C}_0([0,T])$. \vspace{3mm}
\begin{itemize}
    \item Si $H>\frac{1}{2}$, alors $\int_0^t{K_H(t,s)\theta_\epsilon(s)ds}$ converge en loi quand
$\epsilon >0$ vers $B_t^H$.
    \item Si $H<\frac{1}{2}$, on a le même résultat de convergence en loi
mais avec des conditions supplémentaires  qu'on notera dans toute
la suite (CS). Pour plus de détail sur les conditions de type
(CS), on se réfère aux travaux de \cite{B.J.T}. Par exemple, quand
on approche le bruit blanc par le processus:
\begin{equation}
\theta_\epsilon(s)=\sum_{k=1}^{\infty}\xi_kI_{[k-1,k)}(\frac{s}{\epsilon^2}),
\end{equation}
où $\{\xi_{k}, k\geqslant 1\}$ est une suite de variables
aléatoires réelles indépendantes et de même loi, avec
$\mathbb{E}\xi_1=0$, $\text{Var}\xi_{1}^{2}=1$, on suppose que les
variables aléatoires $\xi_k$ ont des moments d'ordre
$m\in\mathbb{N}$ et $m>\frac{1}{H}$.
\end{itemize}
\vspace{3mm} Dans le but d'écrire l'équation de Langevin
généralisée, on traitera séparément les cas où $H$ est supérieur à
$\frac{1}{2}$ et  $H$ est inférieur à $\frac{1}{2}$.\\
\vspace{3mm} \textbf{Premier cas: $H>\frac{1}{2}$} \vspace{3mm}
Considérons l'équation de Langevin (\ref{E.L}):
\begin{equation}
mv'_t=-bv_t+F_t.
\end{equation}
Le processus $\theta_\epsilon(t)$, où $\theta_\epsilon(t)$ est un
bruit blanc quand $\epsilon\rightarrow 0$, (l'existence de ce
processus est dû à (\cite{strook}). On suppose que pour tout
$\epsilon > 0$, on a:
\begin{equation}
F_t+g_{\epsilon(t)}=\theta_\epsilon(t),
\end{equation}
où $g_{\epsilon}$ est une fonction déterministe telle que
$g_{\epsilon}\rightarrow 0$ quand $\epsilon\rightarrow 0$.
Multiplions les deux membres de l'équation (\ref{E.L}) par la
fonction de carré integrable $K_H$ définie par:
\begin{equation}
K_H(t,s)=c_Hs^{\frac{1}{2}-H}\int_s^t(u-s)^{H-\frac{3}{2}}u^{H-\frac{1}{2}}du,
\end{equation}
avec $t>s$ et $c_H=\Big[\frac{H(2H-1)}
{\beta(2-2H,H-\frac{1}{2})}\Big]^{\frac{1}{2}}$. En integrant sur
l'intervalle $[0,t]$, on obtient pour tout $\epsilon > 0$
\begin{equation}
m\int_0^t{K_H(t,s)dv_s}=-b\int_0^t{K_H(t,s)v_sds}+\int_0^t{K_H(t,s)\theta_\epsilon(s)ds}-\int_0^t{K_H(t,s)g_{\epsilon}(s)ds}
\end{equation}
Au passage à la limite $\epsilon\rightarrow 0$,
$\int_0^t{K^H(t,s)\theta_\epsilon(s)ds}$ converge en loi vers
$B_t^H$ dans $\mathcal{C}_0([0,T])$ \cite{B.J.T,CT}. Dans ce cas,
presque sûrement on a l'équation suivante
\begin{equation} \label{eq.frac}
m\int_0^t{K_H(t,s)dV_s}=-b\int_0^t{K_H(t,s)V_sds+\sigma B_t^H}
\end{equation}
L'équation (\ref{eq.frac}) sera utile pour définir l'équation de
Langevin généralisée.\\ \vspace{3mm} \textbf{Deuxième cas:
$H<\frac{1}{2}$} \vspace{3mm} D'une manière analogue en utilisant
le noyau
\begin{equation}
K_H(t,s)=c_H\left[\left(\frac{t}{s}\right)^{H-\frac{1}{2}}(t-s)^{H-\frac{1}{2}}-
   (H-\frac{1}{2})s^{\frac{1}{2}-H}
   \int_s^tu^{H-\frac{3}{2}}(u-s)^{H-\frac{1}{2}}du\right]
\end{equation}
 avec $t>s$ et en choisissant $c_H=\Big[\frac{2H}
{(1-2H)\beta(1-2H,H+\frac{1}{2})}\Big]^{\frac{1}{2}}$, et sous les
conditions de type (CS), on obtient une équation analogue à
(\ref{eq.frac}). \section{Définition de la vitesse fractionnaire}
\textbf{Premier cas: $H>\frac{1}{2}$} \vspace{3mm} Après
approximation du terme $\int_0^t{K_H(t,s)V_s ds}$ par une somme
(au sens de Riemann) et vu que $K_H(t,s) ds$ admet la  même unité
en temps que $t^{H-\frac{1}{2}}$, il est naturel de définir une
vitesse fractionnaire en se basant sur l'intégrale
$\int_0^t{K_H(t,s)V_s ds}$ normalisée par une fonction de type
$\Phi_H(t)=A_H t^{\frac{1}{2}-H}$, où $A_H$ est une constante qui
depend de $H$, dont le rôle sera explicité plus loin. On définit
ainsi la vitesse fractionnaire $V_t^H$ par:
\begin{equation}
\label{VH} V_t^H=V_0+\Phi_H(t)\int_0^t{K_H(t,s)V_s ds}
\end{equation}
Posons $f_H(t)=\int_0^t{K_H(t,s)V_s ds}$ et calculons sa dérivée
par rapport à t:
\begin{eqnarray*}
\frac{df_H(t)}{dt}&=&\lim_{h \rightarrow 0}\left[\frac{1}{h}\left(\int_0^t{K_{H}(t,s)V_sds}-\int_0^{t-h}{K_{H}(t-h,s)V_sds}\right)\right] \\
&=&\int_0^t{\frac{\partial{K_H(t,s)}}{\partial{t}}V_sds}+\lim_{h
\rightarrow
0}\left[\frac{1}{h}\int_{t-h}^{t}{K_H(t,s)V_sds}\right]
\end{eqnarray*}

D'autre part

\begin{equation}
\lim_{h\rightarrow0}\left(\frac{1}{h}\int_{t-h}^{t}{K_H(t,s)V_sds}\right)=0
\end{equation}

On en déduit

\begin{equation}
\frac{df_H(t)}{dt}=\int_0^t{\frac{\partial{K_H(t,s)}}{\partial{t}}V_sds}
\end{equation}

Par suite

\begin{equation}
\frac{dV_t^H}{dt}=\Phi_{H}'(t)f_H(t)+\Phi_H(t)\frac{df_H(t)}{dt}
\end{equation}

Soit donc

\begin{eqnarray*}
\int_0^t{\frac{\partial{K_H(t,s)}}{\partial{t}}V_sds}&=&\frac{1}{\Phi_H(t)}\frac{d(V_t^H-V_0)}{dt}-\frac{\Phi_{H}'(t)}{\Phi_H(t)^2}(V_t^H-V_0)\\
&=&\frac{d(\frac{(V_t^H-V_0)}{\Phi_H(t)})}{dt}
\end{eqnarray*}

Multiplions les deux termes l'équation (\ref{eq.frac}) par
$\Phi_H(t)$, on obtient ainsi:

\begin{equation}
m\,\Phi_H(t)\int_0^t{K_H(t,s)dV_s}=-b\,\Phi_H(t)\int_0^t{K_H(t,s)V_sds+\sigma\,\Phi_H(t)
B_t^H}
\end{equation}

soit

\begin{equation}
m\,\Phi_H(t)\int_0^t{K_H(t,s)\frac{dV_s}{ds}ds}=-b\,(V_t^H-V_0)+\sigma\,\Phi_H(t)
B_t^H
\end{equation}

On introduit la notation suivante:

\textbf{Notation}:

\vspace{3mm} Pour toute fonction
$u(x)=\int_0^x{\frac{\partial{f(x,y)}}{\partial x}g(y)dy}$, on
note:

\begin{equation}
\mathcal{P}^H(u(x))=\int_0^x{f(x,y)\frac{\partial{g(y)}}{\partial
y}dy}
\end{equation}

(Le rôle de $\mathcal{P}^H$ consiste à permuter les dérivées
partielles).

Puisque

\begin{equation}
\frac{d(\frac{(V_t^H-V_0)}{\Phi_H(t)})}{dt}=\int_0^t{\frac{\partial{K_H(t,s)}}{\partial{t}}V_sds}
\end{equation}

il est claire que:

\begin{equation}
\mathcal{P}^H(\frac{d(\frac{(V_t^H-V_0)}{\Phi_H(t)})}{dt})=\int_0^t{K_H(t,s)\frac{dV_s}{ds}ds}
\end{equation}

Finalement, l'equation de Langevin, quand $H>\frac{1}{2}$, s'écrit
sous la forme:
\begin{equation}\label{eq.lan.frac}
m \mathcal{P}^H(\frac{d(\frac{(V_t^H-V_0)}{\Phi_H(t)})}{dt})=-
\frac{b}{\Phi_H(t)}(V_t^H-V_0)+\sigma B_t^H
\end{equation}

\textbf{Deuxième cas: $H<\frac{1}{2}$} \\ Par un raisonnement
analogue, en considérant le noyaux de carré integrable $K_H$
correspondant et sous les conditions de type (CS), on obtient
l'équation (\ref{eq.lan.frac}). \vspace{3mm} En résume, l'équation
généralisée de Langevin sur l'intervalle $[0,T]$ s'écrit:
\vspace{3mm} Pour $H\neq \frac{1}{2}$
\begin{equation}
m \mathcal{P}^H(\frac{d(\frac{(V_t^H-V_0)}{\Phi_H(t)})}{dt})=-
\frac{b}{\Phi_H(t)}(V_t^H-V_0)+\sigma B_t^H
\end{equation}
Pour  $H=\frac{1}{2}$, l'équation de Langevin est donnée par:
\begin{equation}
m dV_t=-b V_tdt+\sigma dB_t
\end{equation}
\begin{remark}
En s'appuyant sur l'équation (\ref{esperance}), et en choisissant
les noyaux correspondants avec les conditions correspondantes, on
a:
\begin{equation}
\mathbb{E}V_t^H=
\mathbb{E}(V_0)(1+\Phi_H(t)\int_0^t{K_H(t,s)e^{-\frac{b}{m}s} ds})
\end{equation}
\end{remark}
\section{Indice de Hurst et l'analyse R/S}
L'indice de Hurst peut être calculer en utilisant la méthode de
l'analyse R/S \cite{RS}. Pour une série temporelle, $X=X_1, X_2,
\dots X_n$, la méthode R/S est comme suit:
\begin{enumerate}
\item On calcule la valeur moyenne $m$:
\begin{equation}
m=\frac{1}{n}\sum_{i=1}^nX_i
\end{equation}
\item On calcule la série Y de l'ajustement moyen:
\begin{equation}
Y_t=X_t-m,\quad t=1,2,\dots,n
\end{equation}
\item On calcule la série Z de la déviation cumulative:
\begin{equation}
Z_t=\sum_{i=1}^tY_i,\quad t=1,2,\dots,n \}
\end{equation}
\item On calcule la série des étendues R:
\begin{equation}
R_t=\max(Z_1,Z_2,\dots,Z_t)-\min(Z_1,Z_2,\dots,Z_t),\quad
t=1,2,\dots,n
\end{equation}
\item On calcule la série R de la déviation standard:
\begin{equation}
S_t=\sqrt{\frac{1}{t}\sum_{i=1}^t(X_i-\mu)^2},\quad
\mu=\frac{1}{t}\sum_{i=1}^tX_i,\quad t=1,2,\dots,n
\end{equation}
\item On calcule la série R/S des étendues normalisées:
\begin{equation}
(R/S)_t=\frac{R_t}{S_t}, \quad t=1,2,\dots,n
\end{equation}
\end{enumerate}
Hurst \cite{RS} avait montré que lorsque $t$ est assez grand, on
a:
\begin{equation}
(R/S)_t=\lambda\,t^H
\end{equation}
Soit donc:
\begin{equation}
\ln ((R/S)_t)=H \ln(t)+\ln(\lambda)\text{où} \, \lambda > 0
\end{equation}
En faisant une regression simple où la variable à expliquer est
$\ln ((R/S)_t)$ et la variable explicative est $\ln(t)$, la pente
de la droite de regression fournit une estimation de l'exposant H
de \cite{RS}. Bien évidemment, Dans la littérature, d'autres
formes de cette méthodes existent et pour cela on peut se référer
au travail de  \cite{RS}. Une estimation de la moyenne de
l'exposant H a été donnée par  \cite{FL}.
\section{Estimation de $A_H$}
Reprenons l'expression de la vitesse fractionnaire $V_t^H$ quand
$H\neq \frac{1}{2}$.
\begin{equation}
V_t^H=V_0+\Phi_H(t)\int_0^t{K_H(t,s)V_s ds},
\end{equation}
où
\begin{equation}
\Phi_H(t)= A_Ht^{\frac{1}{2}-H}.
\end{equation}
On a donc:
\begin{equation}
A_H=\frac{t^{H-\frac{1}{2}}(V_t^H-V_0)}{\int_0^t{K_H(t,s)V_s ds}}
\end{equation}

Après estimation de $H$, on utilise le noyau correspondant $K_H$,
puis on estime $A_H$, de la manière suivante:

On définit sur l'intervalle [0,t], une subdivision
$0<t_1<t_2<\dots<t_n=t$. On choisira $t_0=0$ et $t_1$ très proche
de $0$. Dans ce cas on estime $A_H$ par:

$$\widehat{A_H}=\frac{1}{n}\sum_{i=1}^{n}\frac{t_{i}^{H-\frac{1}{2}}(\widehat{V_{t_i}^H}-V_0)}{\int_0^{t_i}{K_H(t_i,s)V_s ds}}$$

Où $\widehat{V_{t_i}^H}$ est la vitesse fractionnaire estimée à
l'instant $t_i$.

Le rôle de $A_H$ consiste à mieux approcher les valeurs
expérimentales des valeurs théoriques.

\section{Utilisation du modèle}

On considère la subdivision précédente de l'intervalle [0,t], et
on mesure expérimenta\-lement les vitesses fractionnaires
$\widehat{V_{t_i}^H}$, à tout instant $(t_i)_{i=1}^n$. En
utilisant la méthode R/S, on estime le paramètre de Hurst H. Avec
la valeur de H estimée, et en considérant le noyau correspondant,
on estime $A_H$. En revenons à l'équation (\ref{VH}), et en
remplaçant $H$ et $A_H$ par leurs valeurs estimées, on obtient une
expression de la vitesse du phénomène physique considéré. Bien
évidemment, la valeur de $H$, caractérise la série chaotique
correspondant au phénomène physique étudié.
\section{Simulation de la vitesse fractionnaire}
Une simulation a été faite pour certaines valeurs de $H$ quand
$H>\frac{1}{2}$ et  $H<\frac{1}{2}$ pour une valeur fixée de $A_H$
voir figures .
\section{Conclusion}Les mouvements browniens fractionnaires (mBf) modélisent de très
nombreux phéno\-mènes auto-similaires, ou du moins localement
auto-similaires: fluctuations boursières en mathématiques
financières, trafic sur Internet, textures en analyse d'images,
approches de la turbulence en physique des fluides et la liste est
longue. Un domaine  en pleine expansion est celui des équations
différentielles stochastiques régies ou guidées par un mouvement
brownien fractionnaire. Les expérimentateurs cherchent souvent à
disposer de modèles leur fournissant des ensembles de fonctions
génériques qu'ils pourront mieux comparer à leurs observations
qu'un exemple particulier de fonction. Toutes ces raisons
expliquent sans doute la popularité des modèles où l'irrégularité
se conjugue avec l'aléatoire. On fait alors appel au probabiliste
pour construire des processus (fonctions dépendant du hasard) tels
que l'on puisse spécifier leur irrégularité sans pour autant
pouvoir prédire la forme particulière de la fonction générée. Pour
illustrer cette manière de penser, et connaissant les propriétés
du mBf, il apparaît comme un bon processus pour réaliser une telle
modélisation.

\begin{figure}[!h]
\hspace{-1cm} \centering
\mbox{\epsfig{file=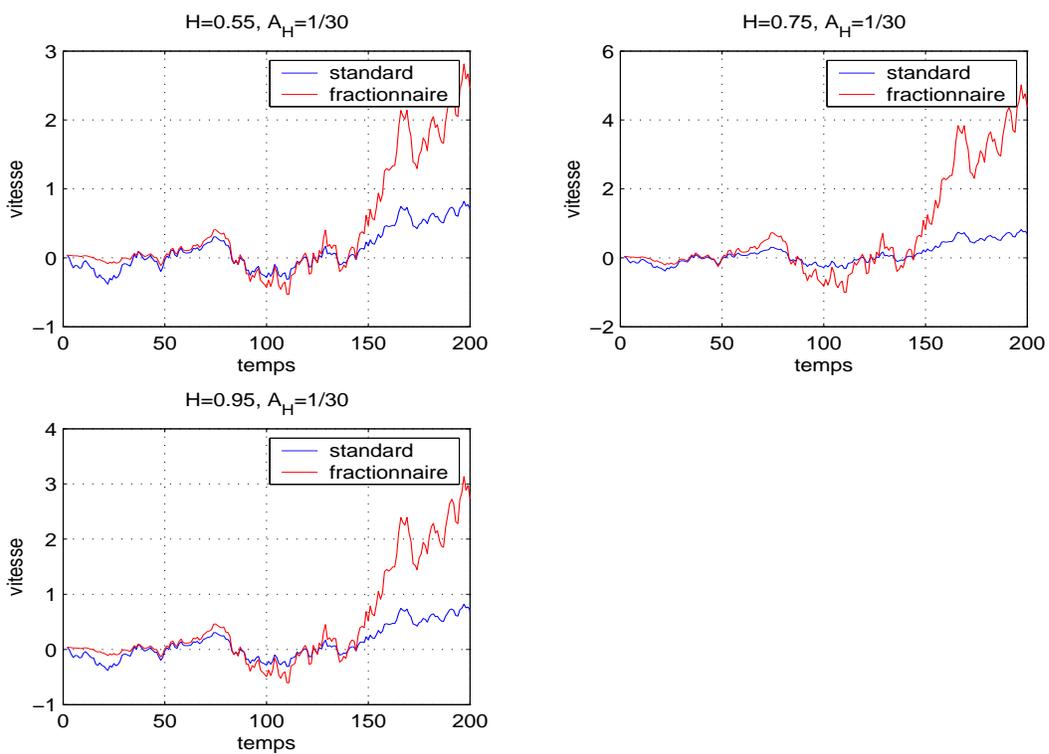,width=14cm,height=10cm}}
\caption{Simulation des vitesses standard et fractionnaires pour
$H>\frac{1}{2}$.} \label{fig:Hsup}
\end{figure}

\begin{figure}[!h]
\hspace{-1cm} \centering
\mbox{\epsfig{file=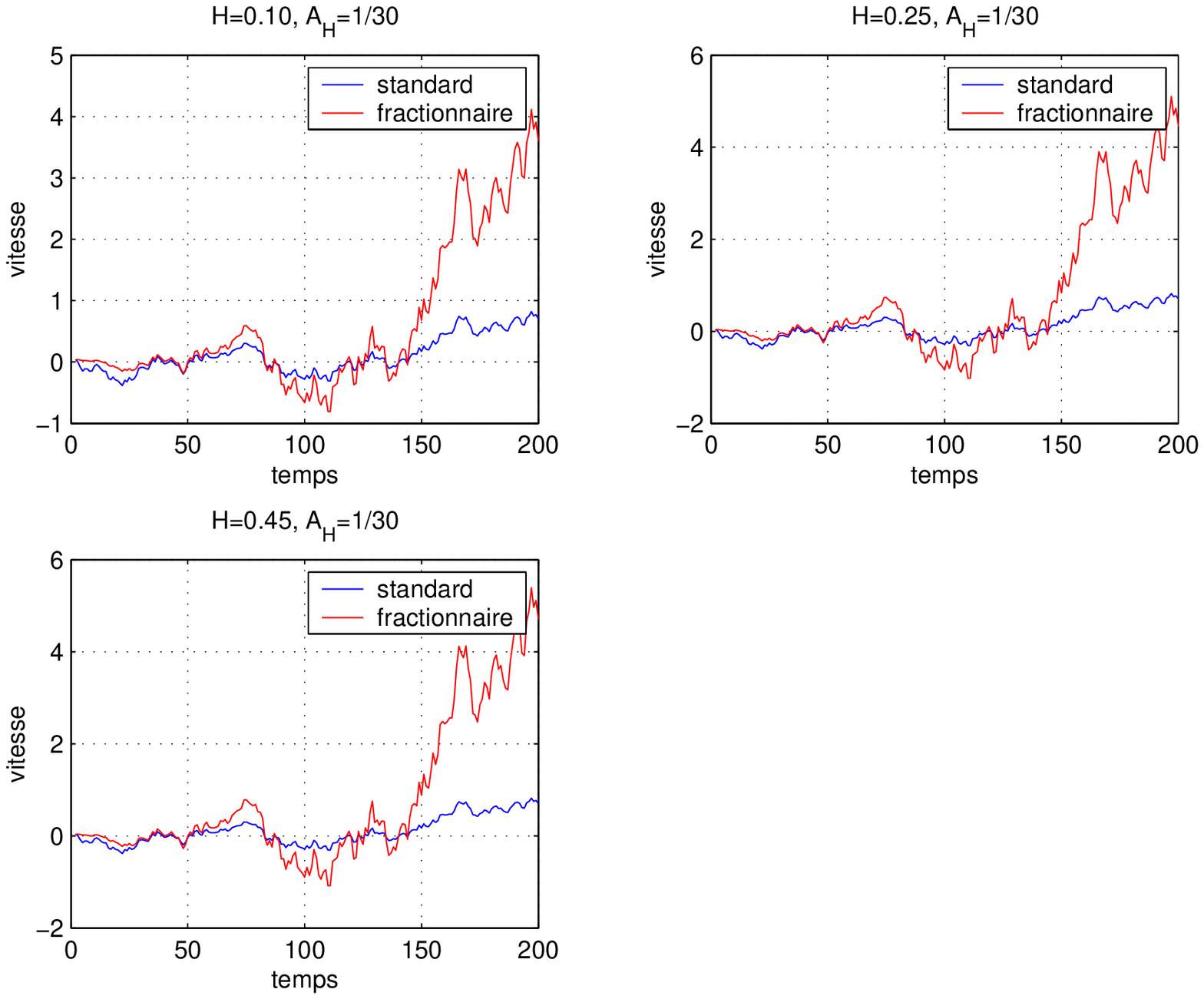,width=14cm,height=10cm}}
\caption{Simulation des vitesses standard et fractionnaires pour
$H<\frac{1}{2}$.} \label{fig:Hinf}
\end{figure}
~~\\

%\bibliographystyle{natbib}
%\bibliography{toufik}

\end{document}